\documentstyle[prb,aps,epsf,multicol]{revtex}

\begin{document}

\title{\Large{\bf Comment on ``Magnetoviscosity and relaxation in ferrofluids''}}
\author{Mark I. Shliomis}
\address{Department of Mechanical Engineering, Ben-Gurion University of the Negev,\\
P.O.B. 653, Beer-Sheva 84105, Israel}
\date{\today}
\maketitle

\draft
\tighten

\begin{abstract}
It is shown and discussed how the conventional system of
hydrodynamic equations for ferrofluids was derived. The set
consists of the equation of fluid motion, the Maxwell equations,
and the magnetization equation. The latter was recently revised by
Felderhof [Phys. Rev. E {\bf 62}, 3848 (2000)]. His
phenomenological magnetization equation looks rather like
corresponding Shliomis' equation, but leads to wrong consequences
for the dependence of ferrofluid viscosity and magnetization relaxation
time on magnetic field.

\end{abstract}
\pacs{PACS number(s): 47.65.+a, 75.50.Mm, 47.15.-x, 83.85.Jn}

\begin{multicols}{2}
\narrowtext
\section{INTRODUCTION. EQUATION OF FLUID MOTION}

In a recent paper [1], Felderhof made an attempt to revise the
conventional hydrodynamic equations for ferrofluids. He proposed
some modification in the equation of ferrofluid magnetization. A
complete set of ferrohydrodynamic equations was worded for the
first time in our paper [2] almost 30 years ago. In the Comment,
we will consider the object of Felderhof's criticism, analyze his
proposition, and explain why is it wrong.

The main peculiarity of ferrofluids is a specific relation between
the magnetic and rotational degrees of freedom of suspended
magnetic grains of which the fluids are composed. Therefore the
concept of {\it internal rotation} first applied to ferrofluids in
[2] has proved to be very fruitful. The model [2] takes into
account that the volume density of the angular momentum of
ferrofluids consists of both the visible (``orbital'') and the
internal (``spin") parts. The former, ${\bf L}=\rho\,({\bf r\times
v})$, is associated with the translational motion of magnetic
grains and molecules of the solvent. The latter, ${\bf S}$, is
caused by the rotation of the grains themselves and should be
treated as an independent variable along with the fluid velocity
${\bf v}$, density $\rho $, and pressure $p$. However, an
appropriate thermodynamic coordinate is the difference ${\bf
S}-I{\bf \Omega }$ where ${\bf \Omega}
={\textstyle{\slantfrac{1}{2}}}\,\rm {curl}\thinspace{\bf v}$ is
the local angular velocity of the fluid and $I$ means the volume
density of the particles moment of inertia. For a suspension of
spherical particles $I=\rho _{s}\phi d^{2}/10$ where $\phi $ is
the volume fraction of the dispersed phase, $\rho _{s}$ the
particles material density, and $d$ the mean particle diameter. In
this case it is convenient to set ${\bf S} = I{\bbox{\omega
}}_{p}$ where ${\bbox{\omega}}_{p}$ is the macroscopic (i.e.,
averaged over physically small volume) angular velocity of the
particles. Any deviation of ${\bbox{\omega}}_{p}$ from ${\bf
\Omega}$ gives rise to dissipation processes due to redistribution
of angular momentum between ${\bf L}$ and ${\bf S}$ forms. (The
angular momentum conservation law refers, sure, to the total
angular momentum ${\bf L+S}\,$). These processes contribute the
stress tensor $\sigma_{ik}$. For an ordinary (nonmagnetic)
suspension the tensor has been derived by the methods of
irreversible thermodynamics in [3]:
\begin{equation}
\sigma _{ik}=-p\delta _{ik}+\eta \left( \frac{\partial v_{i}}{\partial x_{k}}%
+\frac{\partial v_{k}}{\partial x_{i}}\right) +\frac{1}{2\tau _{s}}%
(S_{ik}-I\Omega _{ik}), \eqnum{1}
\end{equation}
where
\[
\ S_{ik}=\epsilon _{ikl}S_{l}\,,\ \ \ \  \Omega _{ik}=(\partial v_{k}/
\partial x_{i}-\partial v_{i}/\partial x_{k})/2=\epsilon _{ikl}\Omega_{l}\,,
\]
and $\epsilon _{ikl}$ stands for antisymmetric unit tensor. Apart from the
viscosity $\eta\,$, Eq. (1) contains once more kinetic coefficient: the spin
relaxation time $\tau _{s}=I/6\eta \phi =\rho _{s}d^{2}/60\eta \,$. For
$d=10\,\rm nm$ and $\eta =10^{-2}\,\rm Ps$ this formula gives
$\tau_{s}\sim 10^{-11}\,\rm s\,(!)$. Thus, the difference
${\bbox{\omega }}_{p}-{\bf \Omega }$ instantly decays whereupon the
hydrodynamic description is reduced to the common set of hydrodynamic
equations. Ferrofluids, however, give us an opportunity to maintain this
difference by an extraneous magnetic torque which acts directly upon the
particles rotation:
\begin{equation}
6\eta \phi ({\bbox{\omega}}_{p}-{\bf \Omega )=M\!\times H}\,. \eqnum{2}
\end{equation}
Here ${\bf H}$ is the magnetic field within the fluid and ${\bf M}$ is the
ferrofluid magnetization. At the {\it equilibrium} in a stationary field,
${\bf M}$ is described well by the Langevin formula
\begin{equation}
{\bf M}_{0}=nmL(\xi )\frac{{\bf H}}{H}\,,\;\; \xi
=\frac{mH}{k_{B}T}\,, \;\; L(\xi )=\coth \xi -\xi ^{-1}, \eqnum{3}
\end{equation}
where $m$ is the magnetic moment of a single particle and $n$ the
number density of the particles. Eliminating the last term in (1)
with the aid of the torque balance equation (2) and including in
$\sigma _{ik}$ the Maxwell tensor of magnetic field, one gets
[2,4]
\begin{eqnarray}
\sigma _{ik}=-p\delta _{ik}+\eta \left( \frac{\partial v_{i}}{\partial x_{k}}%
+\frac{\partial v_{k}}{\partial x_{i}}\right) +\frac{1}{2}(M_{i}H_{k}-%
M_{k}H_{i}) \nonumber \\
+\frac{1}{4\pi }(H_{i}B_{k}-\frac{1}{2}H^{2}\delta_{ik})\,. \eqnum{4}
\end{eqnarray}
On substitution $B_{k}=H_{k}+4\pi M_{k}$ in this tensor, we
are convinced of its {\it symmetry}. Equation (4) and the momentum
conservation law
\[
\rho \frac{dv_{i}}{dt}=\frac{\partial\sigma_{ik}}{\partial x_{k}}\,,
\ \ \ \ \ \ \frac{d}{dt}=\frac{\partial }{\partial t}+({\bf v \nabla })\,,
\]
determine the equation of ferrofluid motion
\begin{equation}
\rho \frac{d{\bf v}}{dt}=-{\bf \nabla }p+\eta \nabla ^{2}{\bf v+(M\nabla )H+}%
{\textstyle{\slantfrac{1}{2}}}\,\text{curl}\,({\bf M\!\times\!H)}\,. \eqnum{5}
\end{equation}
In the calculation of the divergence of the stress tensor we have used the
equations
\begin{equation}
\text{div\thinspace }{\bf v}=0\text{ , \ \ \ \ \ \ curl\thinspace }{\bf H}=0%
\text{ , \ \ \ \ \ \ \ div\thinspace }{\bf B}=0\text{ ,}  \eqnum{6}
\end{equation}
i.e., the ferrofluid is considered to be incompressible and non-conducting.

The system of equations (5)-(6) is still not complete since it
does not determine the ferrofluid magnetization. The latter
influences the fluid motion (see (5)) and depends itself on the
motion as well. There are two basic ways to derive the missing
magnetization equation. Both the ways have been proposed by the
author with co-workers [2,5] and discussed in reviews [4,6,7].

\section{PHENOMENOLOGICAL MAGNETIZATION EQUATION}

Originally the magnetization equation has been derived
phenomenologically [2] as a modification of the Debye relaxation
equation [8]. To get the generalized equation, one should
introduce a local reference frame $\Sigma ^{\prime }$, in which
the suspended particles are quiescent {\it on the average}, i.e.,
${\bbox{\omega}}_{p}^{\prime }=0\,$. It is natural to assume
that the magnetization relaxation is described in the system by the
simplest Debye-like equation
\begin{equation}
\frac{d^{\prime}{\bf M}}{dt}=-\frac{1}{\tau}\,({\bf M-M}_{0}) \eqnum{7}
\end{equation}
with ${\bf M}_{0}$ from (3). Other words, it assumes that any
deviation (either in direction or magnitude) of ${\bf M}$ from its
equilibrium value ${\bf M}_{0}$ decays according to the simple
exponential law $({\bf M-M}_{0})\sim \exp (-t/\tau)$ . Here
$\tau=3\eta V/k_{B}T$ stands for the Brownian time of rotational
particle diffusion ($V=\pi d^{3}/6)\,$ since the particles are
assumed to be rigid magnetic dipoles whose reorientation is
possible only with rotation of the particles themselves. The frame
of reference $\Sigma ^{\prime }$ rotates with respect to the fixed
(``laboratory'') system $\Sigma $ with the angular velocity
${\bbox{\omega}}_{p}\,$. The rates of change of any vector ${\bf
A}$ in systems $\Sigma $ and $\Sigma ^{\prime }$ are related by
the kinematic expression
\begin{equation}
\frac{d{\bf A}}{dt}={\bbox{\omega}}_{p}\!\times\!{\bf A}+
\frac{d^{\prime }{\bf A}}{dt}\text{ .}\eqnum{8}
\end{equation}
Substituting here $\bf A= M\,$,  ${\bbox{\omega}}_{p}$ from (2), and
$d^{\prime }{\bf M/}dt$ from (7), we obtain the equation sought:
\begin{equation}
\frac{d{\bf M}}{dt}={\bf\Omega\!\times\!M-}\frac{1}{\tau}\,({\bf M-M}_{0})%
-\frac{1}{6\eta \phi }\,{\bf M\!\times\!(M\!\times H)} \eqnum{9}
\end{equation}
(Shliomis, 1972). The last (relaxation) term in this equation
describes a process of approach of the vector ${\bf M}$ to its
equilibrium orientation without change of the length of this
vector. Equations (5), (6) and (9) constitute the complete set of
conventional ferrohydrodynamic equations.

Let us compare Eq. (9) with the equation [1]
\begin{equation}
\frac{d{\bf M}}{dt}={\bf \Omega\!\times\!M-}\gamma_{H}({\bf H}_{l}
{\bf -H})-\frac{1}{6\eta \phi }\,{\bf M\!\times\!(M\!\times\!H)}
\eqnum{10}
\end{equation}
(Felderhof, 2000). Here $\gamma _{H}$ is a positive
phenomenological constant and ``${\bf H}_{l}={\bf M}C(M)$ is
expressed in terms of the local magnetization'' [1], that requires
an introduction of a very inconvenient notation
$C(M)=(k_{B}T/mM)L^{-1}(M/nm)$, where $L^{-1}(x)$ means the
function inverse to Langevin function (3). On the face of it, Eq.
(10) is only much less convenient than (9) since the left- and
right-hand sides of (10) contain {\it different} relaxing values:
${\bf M}$ and ${\bf H}_{l}({\bf M)}$. Felderhof claims however
that his ``relaxation equation was {\it derived} from irreversible
thermodynamics (IT), and differs from that {\it postulated} by
Shliomis. The two relaxation equations lead to a different
dependence of viscosity on magnetic field." Let us consider both
these statements.

From the point of view of IT, the relaxation term ${\bf (M}_0-{\bf
M})/\tau $ in (9) is neither more nor less ``postulated" than the
term $\gamma_H({\bf H}-{\bf H}_l)$ in (10). It is worth to remind,
there are two methods of IT, and the both was proposed by Landau.
The first of two, L1, based on conservation laws and the condition
of positive entropy production, was applied for the first time to
the building of hydrodynamics of helium [9], then -- hydrodynamics
of fluids with internal rotation [3], liquid paramagnets [10], and
some other liquids. Relaxation equation (10) has been also derived
by the method. One has to be skilful enough to use this cumbrous
method because it does not lead to the one and only form of sought
equations. In this sense the second method, L2, is much more
definite, simple, and direct than L1. It was first applied just to
the description of relaxation of the order parameter in a
non-equilibrium system [11]. An equilibrium value of the parameter
(${\bf M}_0 $ in our case) corresponds to the minimum of an
appropriate thermodynamic potential $\Phi $ (usually the Gibbs or
Helmholtz free energy) depending on the magnetization ${\bf M}$
and other thermodynamic variables. Thus, at the equilibrium
$\partial\Phi/\partial {\bf M}=0 $. Out of equilibrium this
condition is not satisfied, so the relaxation process occurs: $\bf
M $ changes in time approaching ${\bf M}_0 $. For small deviations
from equilibrium, the derivative $\partial \Phi/\partial {\bf M}$
and the relaxation rate $d{\bf M}/dt $ are small. The relation
between the two derivatives in the Landau theory is reduced to
simple proportionality:
\begin{equation}
\frac{d{\bf M}}{dt}=-\gamma\,\frac{\partial \Phi}{\partial {\bf
M}}\eqnum{11}
\end{equation}
with a constant coefficient $\gamma>0$. Hence we have
\begin{equation}
\frac{d\Phi}{dt}=\frac{\partial \Phi}{\partial {\bf
M}}\,\frac{d{\bf M}}{dt} =-\gamma \Bigl(\frac{\partial
\Phi}{\partial {\bf M}}\Bigr)^{\!2}<0 \eqnum{12}
\end{equation}
as it should be: when a system moves to equilibrium, its free
energy decreases. In the case of a weakly non-equilibrium state of
the system, one can substitute in (11) and (12) the expansion
\begin{eqnarray}
\frac{\partial \Phi}{\partial {\bf M}}= \Bigl(\frac{\partial \Phi}{\partial
{\bf M}}\Bigr)_{\!0} + \Bigl(\frac{\partial ^2 \Phi}
{\partial {\bf M}^2}\Bigr)_{\!0}({\bf M-M}_{0})+\,...\,, \nonumber
\end{eqnarray}
where subscript 0 marks the point of equilibrium. As the first
derivative in this point is equal to zero and the second one is
positive, Eq. (11) turns into Eq. (7) with
$\tau^{-1}=\gamma(\partial ^2 \Phi/\partial {\bf M}^2)_0$ and Eq.
(12) takes the form
\begin{equation}
\frac{d\Phi}{dt}=-\frac{({\bf M-M}_0)^2}{\gamma\,\tau^2}\,.
\eqnum{13}
\end{equation}
Thus, Eq. (7) and hence Eq. (9) are well corroborated by the
method of IT. Equation (10) does also not conflict with IT.
Nevertheless it is wrong. As we show below, it leads to anomalous
result for ferrofluid viscosity and magnetization relaxation time.
The pitfall of IT is discussed in Sec. IV.

The Einstein formula for viscosity of suspension, $\eta =\eta_{0}(1+2.5\phi)$,
was obtained without taking into account the rotational motion of suspended
particles relative to carrier liquid. If however the particles angular velocity
${\bbox{\omega}}_{p}$ does not coincide with the angular velocity of the fluid
${\bf \Omega }\,$, there arise friction forces which manifest themselves in an
additional (so-called {\it rotational}) viscosity $\eta_{r}\,$. As the
difference ${\bbox{\omega}}_{p}-{\bf \Omega }$ is maintained by the magnetic
torque (see (2)), rotational viscosity turns out to be a function of the
dimensionless field strength $\xi $. In a stationary field, the steady
solution of Eq. (9) yields in the linear order in $\Omega \tau$
\begin{equation}
{\bf M-M}_{0}=\tau _{\perp }({\bf \Omega\!\times\!M_0})\,, \,\,\,\,\,
\tau _{\perp }=\frac{2\tau}{2+\xi L(\xi )}\,,
\eqnum{14}
\end{equation}
where $\tau _{\perp }$ is the relaxation time of the transverse (to the
field) component of the magnetization. For the Poiseuille flow or the
planar Couette flow under the field directed along the flow (i.e.,
${\bf H\perp \Omega }$ ), we find
\begin{equation}
{\bf M\!\times H=-}\tau _{\perp }M_{0}H{\bf \Omega }\,, \eqnum{15}
\end{equation}
while for arbitrary orientation of magnetic field the right-hand
side of the expression should be multiplied by $\sin ^{2}\alpha $
where $\alpha $ is the angle between vectors ${\bf H}$ and
${\bf \Omega }\,$. Let us substitute the magnetic torque (15) in (5).
When the field ${\bf H}$ is homogeneous, the magnetic force
$\bf (M\nabla)H$ in the right-hand side of Eq. (5) vanishes, while
two other terms of the equation may be grouped:
\[
\eta \nabla^{2}{\bf v+}{\textstyle{\slantfrac{1}{2}}}\,\text{curl}\,%
({\bf M\!\times H)}=\left(\eta +{\textstyle{\slantfrac{1}{4}}}\,%
\tau_{\perp }M_{0}H\right)\nabla ^{2}{\bf v}\,.
\]
The quantity added here to the ordinary viscosity should be regarded as
rotational viscosity
\begin{equation}
\eta _{r}={\textstyle{\slantfrac{1}{4}}}\,\tau_{\perp }M_{0}H\,.
\eqnum{16}
\end{equation}
Substituting here $M_{0}$ from (3) and $\tau_{\perp }$ from (14),
we derive the formula [2]
\begin{equation}
\eta_{r}(\xi )=\frac{3}{2}\,\eta \phi \,\frac{\xi L(\xi )}{2+\xi L(\xi )}%
=\frac{3}{2}\,\eta\phi \,\frac{\xi -\tanh\xi}{\xi +\tanh \xi }\,.
\eqnum{17}
\end{equation}
In the absence of magnetic field an individual particle ''rolls'' freely
along corresponding shear surface with angular velocity ${\bbox{\omega}}_{p}$
equal to ${\bf \Omega }$ , so that $\eta _{r}(0)=0\,$. Conversely,
$\eta _{r}(\xi )$ attains its limiting value
\begin{equation}
\eta _{r}(\infty )=3\eta \phi/2  \eqnum{18}
\end{equation}
(the saturation) when {\it rolling} of the particle is replaced by
{\it slipping}: the field of sufficiently large intensity guarantees
constancy of the particle's orientation, not allowing it to twist
with the fluid. Note that the result (18) does {\it not depend} on
a concrete form of the magnetization equation but follows directly
from the equation of fluid motion (5). Actually, in the limit
under consideration ${\bbox{\omega}}_{p}=0\,$, so that Eq. (2)
takes the form ${\bf M\times H=-}6\eta \phi {\bf \Omega }\,$.
Substituting this torque in (5), we immediately arrive at (18).
Indeed, the value was obtained by Hall and Busenberg [12] as early
as 1969 {\it without use} of any magnetization equation. In any
case, however, such an equation must not contradict the saturation
value (18). Our formula (17) does satisfy the limit (18), whereas
Felderhof's equation (10) leads to the quite different result. His
final formula in [1] gives the value
\begin{equation}
\eta _{r}^{F}(\infty )=\frac{3}{2}\,\eta \phi \,%
\frac{(nm)^{2}}{6\eta \phi \gamma _{H}+(nm)^{2}}\,, \eqnum{19}
\end{equation}
which is {\it evidently less} than the correct value (18).
According to [1], $\gamma _{H}=\chi /\tau$ where
$\chi =nm^{2}/3k_{B}T$ is the initial magnetic susceptibility.
Substituting the $\gamma_H$ and $\phi =nV$ in (19) we find
$6\eta \phi \gamma _{H}=2(nm)^{2}/3\,$, afterwards Eq. (19) yields
$\eta_{r}^{F}(\infty )=9\eta \phi /10\,$. Thus the ratio of the
Felderhof's limiting value of viscosity to the correct value (18)
is equal to 0.6. Other words, in the limit $\xi =\infty \,$
Felderhof's equation (10) predicts $\omega_{p}=0.4\,\Omega $ (?!)
instead of $\omega _{p}=0$ as it {\it must be}.

\section{MAGNETIZATION EQUATION DERIVED MICROSCOPICALLY}

Both the above-mentioned phenomenological methods allow to obtain
{\it linear} relaxation terms in hydrodynamic equations [like
$({\bf M}_{0}-{\bf M})/\tau$ in (9)] and corresponding quadratic
terms for the rate of the entropy growth or the free energy
diminution (like that in (13)). It is clear that such terms are
valid only for small departures from equilibrium. Indeed, Eq. (9)
describes well the rotational viscosity for arbitrary intensity of
a stationary magnetic field but small values of $\Omega\tau$ (see,
e.g., a good agreement between McTague's experiment [13] and
Shliomis' theory [2]), or for small dimensionless amplitude $\xi $
or frequency $\omega \tau $ of an alternating magnetic field (see,
e.g., experiments on the {\it negative viscosity} and their
explanation in [14]). Meanwhile, to describe successfully the
negative ferrofluid viscosity at finite values of the parameters,
we did need to use in [15,16] a more precise magnetization
equation. Such a {\it macroscopic} equation should be derived from
the kinetic Fokker--Planck equation which provides the {\it
microscopic} description of particle diffusion in colloids. The
program was realized by Martsenyuk, Raikher and Shliomis [5] soon
after the phenomenological magnetization equation (9) was derived
in [2].

The Fokker--Planck equation for a ferrofluid moving in a field
${\bf H}$ has the form [7]
\begin{equation}
2\tau\frac{\partial W}{\partial t}={\bf \hat{R}}\,{\bf (\,\hat{R}}-
2\tau\,{\bf\Omega}-{\bf e\,\times }\,{\bbox{\xi}}\,)W \,, \eqnum{20}
\end{equation}
where ${\bf e=m}/m$ is the unite vector along a particle magnetic moment,
${\bbox{\xi }}= m{\bf H/}k_{B}T\,$, and ${\bf \hat {R}}={\bf e\times }
\partial/\partial {\bf e}$ is the infinitesimal rotation operator.
Equation (20) determines the orientational distribution function
$W({\bf e},t)$ of particles magnetic moments. The macroscopic
magnetization is determined by the relation
${\bf M}(t)=nm\langle {\bf e\rangle }$ where angular brackets denote
statistical averaging with the distribution function. Multiplying Eq.
(20) by ${\bf e}$ and integrate over the angles, we arrive at the equation
\begin{equation}
2\tau\frac{d\langle {\bf e}\rangle }{dt}=2\tau\,{\bf \Omega
\times} \langle {\bf e}\rangle -2\langle {\bf e}\rangle -\langle
{\bf e\!\times\!(e\!\times {\bbox{\xi }})}\rangle \,,  \eqnum{21}
\end{equation}
which however is not closed. Indeed, along with the first moment
of the distribution function, $\langle {\bf e\rangle }$, Eq. (21)
contains the second moment (the last term in the equation). It is
easy to make sure that the equation for the second moment includes
the third one, and so on, thus there is the infinite chain of
cross-linked equations. Ideally, however, one would like to have
only one equation since only the first moment -- magnetization --
has a clear physical meaning. An original scheme of closure of the
first-moment equation (21), titled the {\it effective field
method,} has been proposed in [5]. Let us explain the fruitful
physical idea.

In equilibrium $({\bf \Omega }=0)$ under a constant magnetic field
the stationary solution of Eq. (20) is the Gibbs distribution
\begin{equation}
W_{0}({\bf e})=\frac{\xi }{4\pi \sinh \xi }\,\exp ({\bbox{\xi }}%
{\bf e})\,.  \eqnum{22}
\end{equation}
An averaging of the vector ${\bf e}$ with function (22) gives
expression (3) for the equilibrium magnetization:
\begin{equation}
{\bf M}_{0}=nmL(\xi )\,{\bbox \xi }/{\xi }\,. \eqnum{23}
\end{equation}
Only in true equilibrium the magnetization is one or another function of the
field. In a non-equilibrium state there is {\it no connection} between
${\bf M}$ and ${\bf H}\,$: any arbitrary magnetization may be created -- in
principle -- even in the absence of the field. Nevertheless, one may consider
any value of ${\bf M}$ as an equilibrium magnetization in a certain -- specially
prepared -- magnetic field. This {\it effective field }${\bf H}_{e}$ is related
to the {\it non-equilibrium} magnetization by the {\it equilibrium} relation:
\begin{equation}
{\bf M}=nmL(\xi _{e})\,{\bbox{\xi}}_{e}/{\xi _{e}}\,. \eqnum{24}
\end{equation}
During the equilibrium settling process, the dimensionless
effective field ${\bbox{\xi}}_{e}=m{\bf H}_e/k_{B}T$ tends to the
true field ${\bbox{\xi}}$, so the magnetization (24) relaxes to
its equilibrium value (23). Comparing (23) and (24), we see that
the latter is obtained by averaging of ${\bf e}$ with the
distribution function
\begin{equation}
W_{e}({\bf e})=\frac{\xi_{e}}{4\pi \sinh \xi_{e}}\,\exp
({\bbox{\xi}}_{e}{\bf e}) \,,  \eqnum{25}
\end{equation}
which differs from the Gibbs distribution (22) by replacement of
the true field by the effective one. Carrying out the averaging in
(21) with the function (25), we find the sought equation [5]
\begin{eqnarray}
\frac{d}{dt}\left[L(\xi_{e})\frac{{\bbox{\xi}}_{e}}{\xi_{e}}\right] =
{\bf \Omega}\!\times\!\left[ L(\xi_{e})\frac{{\bbox{\xi}}_{e}}{\xi_{e}}\right]-
\frac{L(\xi_{e})}{\tau\xi_{e}}\,({\bbox{\xi}}_{e}-{\bbox{\xi}})\nonumber\\
-\frac{\xi_{e}-3L(\xi_{e})}{2\tau \xi _{e}^{3}}\;
{\bbox{\xi}}_{e}\! \times \!({\bbox{\xi}}_{e}\!\times
{\bbox{\xi}})\,. \eqnum{26}
\end{eqnarray}
This equation defines the dependence of the effective field
${\bbox{\xi}}_{e}$ upon time, true field ${\bbox{\xi}}\,$, and the
fluid vorticity ${\bf\Omega}$. Its solution ${\bbox{\xi}}_{e}$
being substituted into (24) determines the magnetization of a
moving fluid. In the case of small departures from equilibrium,
the effective field might be represented as a sum of the true field and
some small correlation: ${\bbox{\xi}}_{e}={\bbox{\xi}}+\bbox{\nu}$. Then
from (23) and (24) in the linear approximation in $\bbox{\nu}$ we get
\begin{eqnarray}
{\bf M-M}_0=nm\bigl[L'(\xi)\,\bbox{\nu}_{\parallel}+L(\xi)\,\bbox{\nu}_{\perp}
\bigr]\,,
\eqnum{27}
\end{eqnarray}
where the components
\begin{eqnarray}
\bbox{\nu}_{\parallel}={\bbox{\xi}(\bbox{\nu\xi})}/{\xi^2}\,,\;\;\;
\;\;\bbox{\nu}_{\perp}={\bbox{\xi}\!\times\!(\bbox{\nu}\!\times
\bbox{\xi})}/{\xi^3}
\nonumber
\end{eqnarray}
are parallel and perpendicular to the true field, respectively. Employing
the relation (27), one can reduce Eq. (26) to the linear magnetization equation
\begin{eqnarray}
\frac{d{\bf M}}{dt}={\bf \Omega \!\times\!
M}-\frac{{\bf H\,[H}\,({\bf M-M}_{0})]}{\tau_{\parallel}H^{2}}
-\frac{\bf H\!\times\!(M\!\times \!H)}{\tau _{\perp }H^{2}}\,,
\eqnum{28}
\end{eqnarray}
where relaxation times of the components of magnetization are
\begin{equation}
\tau _{\parallel }=\frac{d\ln L(\xi )}{d\ln \xi }\,\tau\text{ , \
\ \ \ \ \ \ }\tau _{\perp }=\frac{2L(\xi )}{\xi -L(\xi )}\,\tau\,.
\eqnum{29}
\end{equation}
Substituting $\tau_{\perp }$ from (29) in (16), we obtain [5]
\begin{equation}
\eta _{r}(\xi )=\frac{3}{2}\,\eta \phi \,\frac{\xi L^{2}(\xi )}
{\xi - L(\xi )} \,.  \eqnum{30}
\end{equation}
In the same approximation our phenomenological equation (9) also
takes the form (28) but with other relaxation times:
$\tau_{\parallel}=\tau$ and $\tau_{\perp}$ is defined in (14).
\begin{figure}
\begin{center}
\epsfxsize=8.5cm
\epsfbox{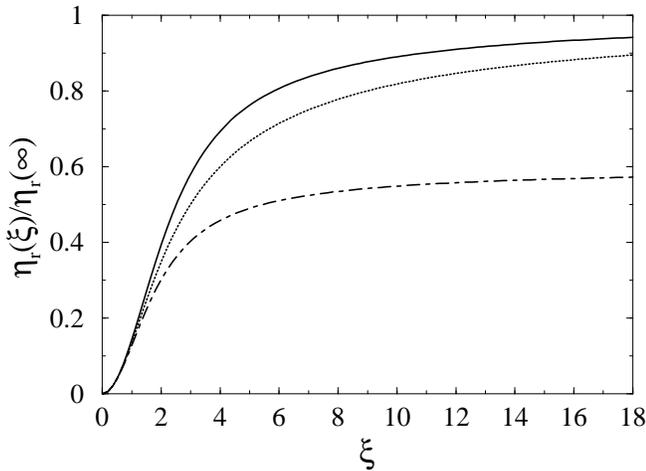}
\end{center}
\vspace*{-0.5cm} \caption{Reduced rotational viscosity
$2\eta_r(\xi)/3\eta\phi$ as a function of $\xi$ given by the
effective field method [5] (Eq. (30), solid curve), by the
phenomenological approach [2] (Eq. (17), dotted curve), and by
Felderhof's approximation [1] (Eq. (32), dash-dotted curve).}
\label{f1}
\end{figure}

Figure 1 shows that though at first sight functions (17) and (30)
do not appear alike, they agree fairly closely in the entire range
of their argument. Both they are in a good agreement with
experimental data of many authors and with computational results
provided by direct numerical integration of the Fokker--Planck
equation in linear approximation in $\Omega \tau$ [17]. Quite
recently Felderhof [18] solved this linearized equation by the
Galerkin method using a large number of trial functions (the
associated Legendre functions). Comparing his ``exact result" for
the rotational viscosity with (30) and (17), he wrote that ``the
result of Martsenyuk, Raikher and Shliomis [5] is quite a good
approximation, but the result of Shliomis [2] deviates up to 17
percent". It is a pity that Felderhof forgot to mention in [18]
his own prediction [1] -- see the bottom curve in Fig. 1 -- which
deviates up to 40 percent!

\section{DISCUSSION}

It is easy to see that Felderhof's ``local field" ${\bf H}_l$ in
(10) is none other than our effective field ${\bf H}_e$ determined
by (24). Hence one can use the relationship (27) for a linear
analysis of Eq. (10). Substituting in (27)
$\bbox{\nu}=\bbox{\xi}_e-\bbox{\xi}=(m/k_BT)({\bf H}_e -{\bf H})$,
we find
\begin{eqnarray}
({\bf H}_e -{\bf H})_{\parallel}=\frac{({\bf M-M}_0)_{\parallel}}{3\chi L'(\xi)}
\,,\;\;\;\;({\bf H}_e)_{\perp}&=&\frac{\xi\,{\bf M}_{\perp}}{3\chi L(\xi)}\,.
\eqnum{31}
\end{eqnarray}
Let us substitute these relations in Felderhof's equation (10) and put there
$\gamma_H=\chi /\tau\,$: with such a choice his equation coincides with (9)
in the limit $\xi\ll 1$ as it should be. Then for arbitrary $\xi$ we obtain
from (10), making use of (31) and (16),
\begin{equation}
\tau_{\perp}^F =
\frac{6L(\xi)\,\tau}{\xi\,[2+3L^2(\xi)]}\,,\;\;\;\;
\eta_r^F=\frac{9}{2}\,\eta\phi\,\frac{L^2(\xi)}{2+3L^2(\xi)}\,.
\eqnum{32}
\end{equation}
Both the relaxation time and viscosity are wrong. Indeed, in a strong field,
$\xi\gg 1\,$, they take the magnitudes $\tau_{\perp}^F=6\tau/5\xi\,$ and
$\eta_r^F=9\eta\phi/10\,$, while it should be
$\tau_{\perp}=2\tau/\xi=6\eta V/mH\,$ and $\eta_r=3\eta\phi/2\,$. The
dependence $\eta_r^F(\xi)$ shown in Fig. 1 strongly differs from two other
curves.

Let us give consideration to the question, why does Felderhof's
equation lead to the anomalous results. We have shown above how
does Debye equation (7) originate from the potential $\Phi(\bf M)$.
One can choose, however, as an independent variable the effective
field and introduce the potential $\tilde{\Phi}({\bf H}_e)$. Then
instead of (11) we obtain in similar fashion
\begin{eqnarray}
\frac{d{\bf H}_e}{dt}=-\tilde{\gamma}\,\frac{\partial
\tilde{\Phi}}{\partial {\bf H}_e}\,,\;\;\;\;\;\;\;\tilde{\gamma}>0.
\nonumber
\end{eqnarray}
Acting further by the L2 method, we arrive at the equation (cf. (7))
\begin{eqnarray}
\frac{d'{\bf H}_e}{dt}=-\frac{1}{\tau}\,({\bf H}_e-{\bf H})\,,
\eqnum{33}
\end{eqnarray}
where we set $\tilde{\gamma}^{-1}=(\partial^2\tilde{\Phi}/
\partial{\bf H}_e^2)\tau $. Under the choice, Eq. (33) turns into (7)
in the low field limit.

Equations (7) and (33) satisfy the principal propositions of the theory
of {\it linear response} according to which the time rate $\dot x$ of
change of a value $x$ at each a moment is determined by the value $x$
at the same moment: $\dot {x}=\dot {x}(x)$. Then, if $x$ weakly deviates
from its equilibrium value $x_0$, one can expand $\dot {x}(x)$ over $x$
and confine oneself to the linear term: $\dot {x}=-\lambda (x-x_0)$,
where $\lambda$ is a positive constant. Thus, $\tau$ in (7) and
(33) should be considered as a {\it constant}. This
inference seems important because the method L1 of irreversible
thermodynamics does not allow {\it in principle} to determine the field
dependence of kinetic coefficients such as $\gamma_H$ in (10). In the
rotating reference frame $\Sigma '$ the equation reads
\begin{eqnarray}
\frac{d'{\bf M}}{dt}=-\gamma_H\,({\bf H}_e-{\bf H})\,.
\nonumber
\end{eqnarray}
In contrast with (7) and (33), this equation relates one value
(${\bf H}_e$) with time rate of change of another one (${\bf M}$).
Therefore, under the nonlinear magnetization law (24), the
coefficient $\gamma_H$ wittingly cannot be constant but represents
a {\it unknown function} of $\xi\,$.

Equation (33) has been written out in a coordinate system $\Sigma
'$. Reverting to the immobile system $\Sigma $ by the general
formula (8) and eliminating $\bbox{\omega}_p$ with the aid of (2),
we obtain [19]
\begin{equation}
\frac{d{\bf H}_e}{dt}={\bf \Omega}\!\times\!{\bf H}_e-\frac{1}{\tau}\,%
({\bf H}_{e}-{\bf H})-\frac{1}{6\eta \phi }\,{\bf H}_e\!\times\!%
({\bf M}\!\times\!{\bf H}).
\eqnum{34}
\end{equation}
This equation determines together with (24) the magnetization $\bf M$
in an implicit form, effective field ${\bf H}_e$ being the parameter.
In the case of small departures from equilibrium, Eq. (34) can be
linearized with respect to ${\bf H}_e-{\bf H}$ and ${\bf M-M}_0$. Using
relationships (31) and
\begin{eqnarray}
\frac{d{\bf M}}{dt}=\frac{d{\bf M}}{d{\bf H}_e}\frac{d{\bf H}_e}{dt}=
3\chi\!\left[L'(\xi)\,\frac{d({\bf H}_e)_{\parallel}}{dt}+
\frac{L(\xi)}{\xi}\,\frac{d({\bf H}_e)_{\perp}}{dt}\right]
\nonumber
\end{eqnarray}
we turn to Eq. (28) with $\tau_{\parallel}=\tau$ and $\tau_{\perp}$ from
(14). Thus, in linear approximation Eqs. (9) and (34) {\it coincide} with
each other. As the result, both the equations yield {\it the same}
relationship (17) for the rotational viscosity of ferrofluids.

It is worth to note, if Felderhof had used ${\bf H}_e$ as an
independent variable {\it correctly}, he would have arrived at Eq.
(34). However, he has missed the opportunity.

\section{CONCLUSION}

Thus, Shliomis' theory consists of hydrodynamic and Maxwell
equations (5)-(6) plus a magnetization equation. There are three
kinds of the latter: (9), (26), and (34). It is well-established
that Eq. (26) derived by the effective field method from the
Fokker--Planck equation yields quite accurate results for real
ferrofluids. Indeed, a direct numerical simulation of the magnetic
moment Brownian dynamics performed by Cebers [20,21] in the middle
of 80th has indicated that Eq. (26) describes perfectly the fluid
magnetization in a wide range of parameters $\xi $ and
$\Omega\tau$. The same conclusion has been made in [6] under
comparing of the solution of (26) with the results of numerical
integration of the non-stationary Fokker--Planck equation (20). At
the same time, the calculations [20,6] have shown that the
phenomenological equation (9) is valid for any field magnitudes
$\xi$ but only small enough fluid vorticities: $\Omega\tau\leq 1$.
Hence Eq. (9) can be recommended to the description of weakly
nonequilibrium situations, as the equation is far simpler for
analysis than Eq. (26). The latter, however, guarantees the
correct quantitative description of magnetization processes even
if deviations from the state of equilibrium are large,
$\Omega\tau\gg 1$, that is when Eq. (9) leads to erroneous
results. Interestingly, our latest calculations have shown that
the new equation for the effective field (34) is free from such a
shortcoming: it is valid even far from equilibrium. Therefore,
taking into account that (34) is nevertheless simpler than (26),
one should recommend Eq. (34) to the most wide applications.

As for the Felderhof's equation (10), it does not stand up to comparison
even with our phenomenological equation (9) to say nothing of the
microscopically derived Eq. (26). We have shown that incorrectly derived
Eq. (10) leads to anomalous results (19) and (32) -- see also Fig. 1 --
and that is why it should be rejected.

\vskip 0.2 truecm

This work was supported by the Meitner--Humboldt Research Award
founded by the Alexander von Humboldt Foundation.

\end{multicols}{2}


\vskip -0.2 truecm

\begin{references}

\bibitem{1} B.U. Felderhof, Phys. Rev. E {\bf 62}, 3848 (2000).

\bibitem{2} M.I. Shliomis, Sov. Phys. JETP {\bf 34}, 1291 (1972).

\bibitem{3} M.I. Shliomis, Sov. Phys. JETP {\bf 24}, 173 (1967).

\bibitem{4} M.I. Shliomis, Sov. Phys. Usp. {\bf 17}, 153 (1974).

\bibitem{5} M.A. Martsenyuk, Yu.L. Raikher, and M.I. Shliomis, Sov. Phys. JETP
{\bf 38}, 413 (1974).

\bibitem{6} M.I. Shliomis, T.P. Lyubimova, and D.V. Lyubimov, Chem. Eng. Comm.
{\bf 67}, 275 (1988).

\bibitem{7} Yu.L. Raikher and M.I. Shliomis, Adv. Chem. Phys. Ser. {\bf 87}, 595
(1994).

\bibitem{8} P. Debye, {\it Polar Molecules} (Dover, New York, 1929).

\bibitem{9} I.M. Khalatnikov, Sov. Phys. JETP {\bf 23}, 8 (1952); I.L. Bekharevich
and I.M. Khalatnikov, ibid. {\bf 40}, 920 (1961).

\bibitem{10} M.I. Shliomis, Sov. Phys. JETP {\bf 26}, 665 (1968).

\bibitem{11} L.D. Landau and I.M. Khalatnikov, Dokl. Akad. Nauk SSSR {\bf 96},
469 (1954).

\bibitem{12} W.F. Hall and S.N. Busenberg, J. Chem. Phys. {\bf 51}, 137 (1969).

\bibitem{13} J.P. McTague, J. Chem. Phys. {\bf 51}, 133 (1969).

\bibitem{14} F. Gazeau, J.-C. Bacri, R. Perzynski, and M.I. Shliomis, Phys. Rev. E
{\bf 56}, 614 (1997).

\bibitem{15} J.-C. Bacri, R. Perzynski, M.I. Shliomis, and G.I. Burde, Phys. Rev.
Lett. {\bf 75}, 2128 (1995).

\bibitem{16} A. Zeuner, R. Richter, and I. Rehberg, Phys. Rev. E {\bf 58}, 6287
(1998).

\bibitem{17} A.C. Levi, R.F. Hobson, and F.R. McCourt, Canad. J. Phys. {\bf 51},
180 (1973).

\bibitem{18} B.U. Felderhof, Magnetohydrodynamics {\bf 36}, 396 (2000).

\bibitem{19} The effective field equation (34) derived from an irreversible
thermodynamic treatment is published in the Comment for the first
time.

\bibitem{20} A.O. Cebers, Magnetohydrodynamics {\bf 20}, 343 (1984); {\bf 21}, 357 (1985).

\bibitem{21} E. Blums, A. Cebers, and M. Maiorov, {\it Magnetic Fluids} (W. de
Gruyter, Berlin, New York, 1997).




\end{references}
\end{document}